\documentclass[prd,twocolumn,nofootinbib]{revtex4}
\usepackage{graphicx}
\usepackage{amsmath}
\usepackage{mathtools}
\usepackage{blkarray, bigstrut} %
\usepackage{units}
\usepackage{physics}
\usepackage{csquotes}
\usepackage[american]{babel}
\usepackage{hyperref}
\DeclareQuoteStyle[american]{english}
  {\textquotedblleft}
  [\textquotedblleft]
  {\textquotedblright}
  [0.05em]
  {\textquoteleft}
  {\textquoteright}
\MakeOuterQuote{"}

\begin{document}

\title{Closed Timelike Curves and "Effective" Superluminal Travel with Naked Line Singularities}

\author{Caroline Mallary} 
\affiliation{Department of Physics, University
  of Massachusetts, Dartmouth, MA 02747}

\author{Gaurav Khanna} 
\affiliation{Department of Physics, University
  of Massachusetts, Dartmouth, MA 02747}

\author{Richard H.~Price} \affiliation{Department of Physics, MIT, 77 Massachusetts Ave., Cambridge, MA 02139}
\affiliation{Department of Physics, University  of Massachusetts, Dartmouth, MA 02747}

\begin{abstract}
We examine closed timelike curves (CTCs) and "effective" superluminal travel in a  spacetime containing naked line singularities, which we call "wires".  Each wire may be straight-line singularity or a ring singularity.  The Weak Energy Condition (WEC) is preserved in all well-defined regions of the spacetime. (The singularities themselves are not well-defined, so the WEC is undefined there, but it is never explicitly violated.)  Parallel to the wire, "effective" superluminal travel is possible, in that the wire may be used as a shortcut between distant regions of spacetime. Our purpose in presenting the superluminal aspects of the wire is to dispel the commonly held view that explicit WEC violation is necessary for effective superluminal travel, whereas in truth the strictures against superluminal travel are more complicated.   We also demonstrate how the existence of such "wires" could create CTCs.  We present a model spacetime which contains two wires which are free to move relative to each other.  This spacetime is asymptotically flat:  It becomes a Minkowski spacetime a finite distance away from each of the wires.  The CTCs under investigation do not need to enter the wires' singularities, and can be confined to regions that are weak-field:  This means that if these wires were physically possible, they would present causality problems even in nonsingular, energetically realistic regions of the spacetime.  We conclude that the Weak Energy Condition alone is not sufficient to prevent superluminal travel in asymptotically flat spacetimes.  

\end{abstract}

\maketitle

\begin{section}{Introduction} \label{sec:intro}

\noindent{\em{Background.}}
General relativity has a `time machine' problem:  Namely, it cannot rule out the possibility of creating time machines. It has been known for many decades that general relativity (GR) allows for spacetimes where time travel is possible \cite{stockum}.  Spacetimes which allow for time travel contain paths which loop back on themselves in time:  By following such a path, an observer could return to his own past.  These looped paths are more properly known as Closed Timelike Curves (CTCs).  For an overview of CTC solutions, see writings by Lobo \cite{lobo} or Thorne \cite{thorneCTC}. \par

Time machines are problematic, both logically and theoretically.  Logical objections to time machines include the famous grandfather paradox and all its science-fictional variations:  These objections are outside the scope of this paper, but are discussed by Lobo \cite{lobo} and Visser \cite{visserBook}.  Theoretically, time machines are problematic because they imply an `incompleteness' in GR:  The evolution of a spacetime with CTCs has no clear, consistent causal structure that can be described by GR itself, or by any other well-accepted theory.

It is also known that `time machine' spacetimes may arise from attempts to devise a spacetime where "effective" superluminal travel is possible, e.g. spacetimes containing traversible wormholes \cite{wormhole}. The problem of time machines is therefore linked to the problem of superluminal travel. 

It is generally assumed that certain `realism' conditions will conspire to prevent CTCs and superluminal travel schemes.  These realism conditions are not inherent to GR, but imposing these conditions is considered much more reasonable than accommodating new notions of causality.  Broadly speaking, these realism conditions fall into 3 categories:

\begin{enumerate}  
\item  The requirement that a spacetime have an `asymptotically flat' geometry.  This means that the spacetime should approach a flat, Minkowski spacetime far from some compact region of interest.  This is important for any spacetime where we are not studying cosmological effects. 

\item  The requirement that the spacetime contain matter which has the properties of realistic matter, as we understand it.  Here, `realistic matter' means matter which meets various energy conditions relating density, pressure and momentum in the stress-energy tensor.  For a detailed description of the energy conditions, see the primer by Curiel \cite{curiel}.

\item The requirement that a spacetime containing CTCs be able to evolve from initial data with no objectionable features.  This requirement is the basis of Hawking's Chronology Protection Conjecture \cite{hawking}.  Ori \cite{oriCTC} breaks down this requirement more specifically as part of his criteria for a physical time machine model.  We will refer to this as the `evolution' requirement. 
\end{enumerate}

Several more minor requirements exist.  For instance, we would like our CTCs to be stable in time; this can be considered part of the `evolution' requirement.  We would also like our spacetime metrics to be smooth to at least the first derivative in every coordinate, and preferably to higher derivatives as well.  This is because the stress-energy tensor of GR contains second derivatives of the spacetime metric, and discontinuities (or `corners') in spacetime are equivalent to infinities in energy density and pressure.  There is an an ambiguous exception to the `smoothness' rule: Singularities are, by definition, not smooth, and according to GR, singularities do exist inside black holes, and do result in a breakdown of causal structure.  The singularities inside rotating black holes are ring singularities.  For most purposes in physics, black hole singularities can be ignored for the simple reason that they are hidden inside a black hole's event horizon.  It is still an open question whether or not `naked' singularities can exist in our universe.  The hypothesis that naked singularities do not exist is known as the Cosmic Censorship Hypothesis \cite{censor}.  If the Cosmic Censorship Hypothesis is correct, then we should never be able to see any discontinuities in spacetime outside an event horizon.  We would therefore require a realistic spacetime to be smooth everywhere, except in regions hidden by an event horizon.\newline

\noindent{\em{Model.}}  In this paper, we present a family of `wire' spacetimes, which allow `effective' superluminal travel, much like traversible wormholes. As a consequence, these spacetimes can also contain CTCs.  All these spacetimes contain one or more extremely dense `wires' of matter, which become singular at their core.  These spacetimes meet most or all of the energy conditions in all well-defined regions.  In particular, they meet the Weak Energy Condition (WEC).  These spacetimes are smooth and asymptotically flat, and their CTCs are not hidden behind event horizons.  We do not attempt to describe how such `wire' spacetimes could evolve from known spacetimes: The evolution requirement is outside the scope of this paper and its goal of demonstrating that explicit WEC-violation is not an inherent feature of superluminal travel schemes.  However, we point out that the mechanism which creates CTCs in our wire spacetimes does not require fine-tuning:  The CTCs exist in an extensive region of the spacetime whenever some fairly broad conditions are met.  We simply need two wire-containing regions to be in proximity to one another, and to have some relative velocity parallel to their lengths.  

The wire-based CTCs are created by a mechanism mathematically similar to the CTCs of the Morris-Thorne Wormhole\cite{wormhole}.  We will discuss the relationship between wire-containing regions and the energy conditions.  
  
The wire spacetimes allow effective superluminal travel and contain CTCs, yet meet the WEC and may have a reasonable, compact shape of singularity (namely, a ring singularity).  We conclude that the Weak Energy Condition alone is not up to the task of forbidding superluminal travel in asymptotically flat spacetimes.\newline

\hfill\newline
\noindent{\em{Fast-Light Regions.}}  For definiteness, we will always be refering to static spacetimes in this section.\par
Our wire spacetimes produce CTCs because they allow for `effective' superluminal travel.  When we say `effective' superluminal travel, we mean that there exist paths in a curved spacetime where light has a coordinate speed $c_{path}$ greater than $c$, where $c$ is the standard speed of light, and $c_{path}$ is measured according to the coordinates of a distant inertial observer to whom space is asymptotically flat (see Figure \ref{fig:fastLightRegion}).  In natural units, we can write this more concisely as:  $c = 1$, $c_{path} > 1$.\par
By extension, this means that massive objects can also have a coordinate speed $v_{path} > 1$ along these paths, provided that their velocity $v_{path}$ is still $v_{path} < c_{path}$.  Two well-known CTC spacetimes with this property are Morris-Thorne wormholes\cite{wormhole} and Gott's cosmic strings\cite{gott}.  Our wires are fundamentally different from Gott's cosmic strings, but they share this effectively superluminal quality.  

Moving forward, we will refer to effectively superluminal regions as `fast-light' regions.

\begin{figure}[ht]
  \centering
  \includegraphics[scale=0.35]{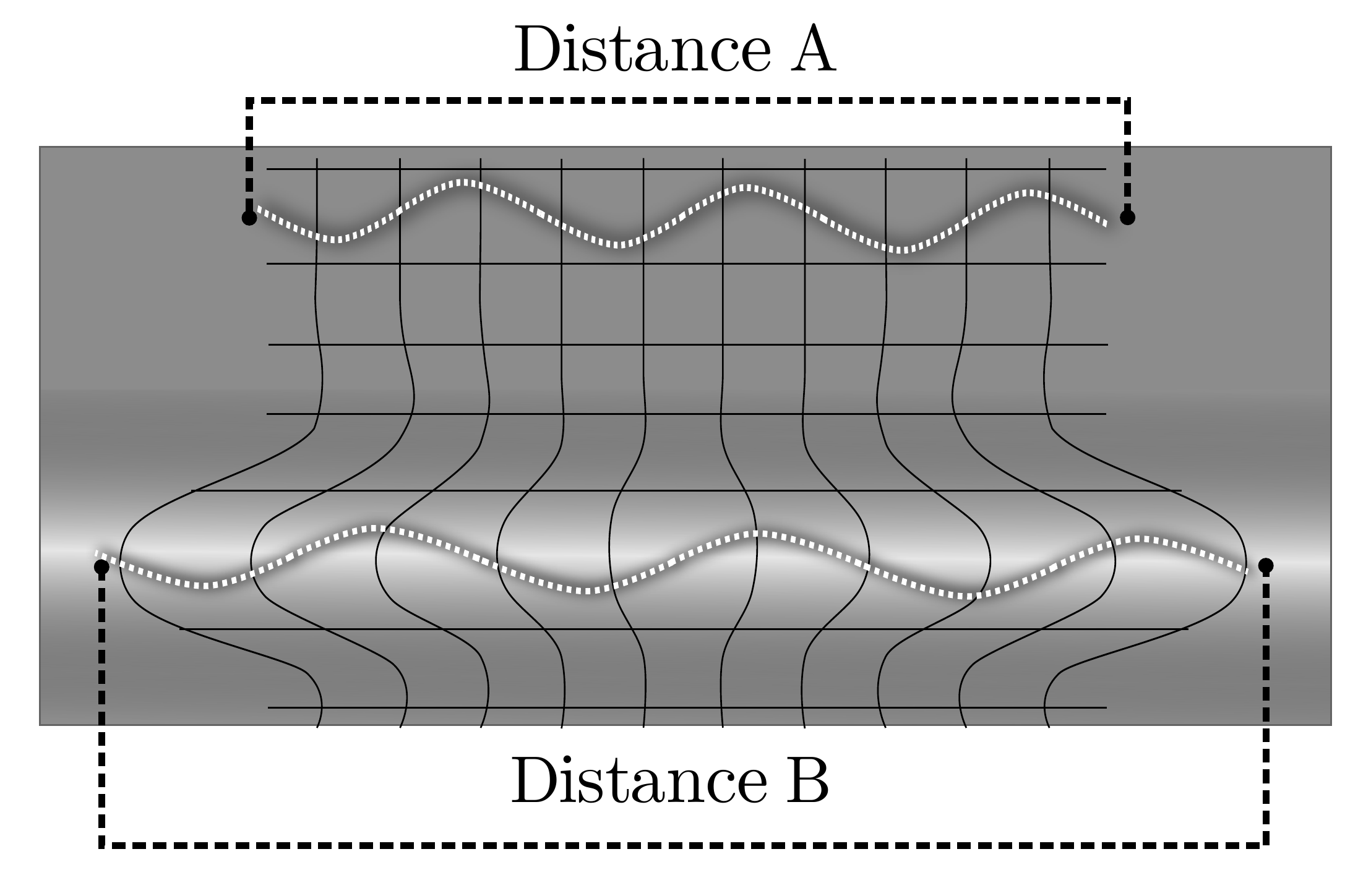}
  \caption{A spacetime whose lower half is a fast-light region. Two identical photons (the sinusoids) travel across the spacetime at different locations.  To these photons, Distance A = Distance B = 3 wavelengths. We see both photons complete their journey in the same amount of time, according to the coordinates of a distant observer.  But according to these same coordinates, Distance B $>$ Distance A.  If the upper photon has velocity $c$, we can infer that the lower photon has velocity $c_{path} = c \times \frac{B}{A} > c$ in the coordinates we are using. To us, then, the lower photon appears to be traveling superluminally. This is why the lower half of the spacetime is a fast-light region.}
  \label{fig:fastLightRegion}
\end{figure} 

Fast-light regions can be formulated in different ways. Here we demonstrate a simple way to formulate a fast-light region using a 
static spacetime. Let us further assume that there exists a special coordinate system in which the metric of this spacetime is diagonal:
\begin{equation} \label{eq:staticMetric} 
ds^2 = \mathbf{g}_{tt}(x)\,dt^2 + \Sigma_i^3\mathbf{g}_{ii}(x)\,(dx^i)^2
\end{equation}
Here, were use natural units, and $t$ is the time coordinate which points in the direction normal to the spacelike hypersurfaces in the spacetime.  The spatial directions $x^i$ must be orthogonal to each other to ensure that our metric is diagonal, but otherwise the $x^i$ axes can be freely chosen.  We would like to measure the coordinate speed of light in some direction $x^{path}$.  For simplicity, let's choose our coordinate system such that $x^{path}$ lies parallel to the $x^3$ axis.  We then have a very simple expression for the coordinate speed of light in the $x^3$ direction:
\begin{equation} \label{eq:c}
c_{path} = \sqrt{  \abs{\frac{ \mathbf{g}_{tt} }{ \mathbf{g}_{33} }}  }
\end{equation}
If $c_{path} > 1$ in natural units, then we have created a fast-light region in the $x^3$ direction. 

For our wire spacetimes, we will be primarily interested in the coordinate speed of light along the length of the wire.  If the wire runs in the $z$ direction, we will have $c_z > 1$.

Visser, et al.\cite{visser}, have argued that `effective' superluminal travel, such as that allowed by fast-light regions, is associated with violations of the null energy condition (NEC).  However, that analysis requires that gravity be weak everywhere, and the authors note that attempts to extend the analysis to the strong-field regime produced unsatisfying results.  The fast-light spacetimes which we will present here all meet the NEC, as well as the more-stringent Weak Energy Condition (WEC).  But our fast-light `wire' spacetimes are strong-field, according to Visser's usage: That is, they cannot be approximated as a perturbation on flat space, so they do not fall under Visser's analysis.

A similar challenge comes from Olum \cite{olum}, and is expanded upon by Lobo \& Crawford \cite{loboCrawford}.  These papers argue that effective superluminal travel requires violation of the WEC.  Again, the singularities in our wire spacetimes seem to exclude them from this constraint:  Here, the analysis requires that there exist some `best', fastest path between two events A \& B.  This is a crucial feature of Olum's definition of `superluminal', but it is difficult to see how a `best' path can be defined in the presence of a naked singularity.  We discuss this further in Section \ref{sec:outro}.

Technically, all energy conditions are undefined at a singularity.  This means that they are neither violated nor not-violated, simply undefined.  In a spacetime which contains naked singularities, we must either 1) accept that the WEC cannot be defined at the singularity itself, or 2) declare that the WEC is automatically unsatisfied in all such spacetimes, since it is not defined everywhere.  We take the first position.  This paper assumes that the WEC can be undefined at a naked singularity while maintaining its relevance outside the singularity.  However, we will show that the WEC is not up to the task of forbidding superluminal travel in certain spacetimes which contain naked singularities.

It is important to note that the CTCs produced by our wire spacetimes need not approach the central singularity at all.  For this reason, our analysis of CTCs in Section \ref{sec:CTCs} makes no mention of singularities.  However, it appears that naked singularities may be necessary for fast-light regions to meet the WEC.  This is suggested by the fact that the wires manifestly \emph{do} meet the WEC, despite proofs which associate superluminal travel with WEC-violation in non-singular spacetimes (\cite{visser},\cite{olum},\cite{loboCrawford}).  A full analysis of the role of the singularity in WEC-preservation is beyond the scope of this work.  We note that we found no way to remove the singular region from our metrics while also preserving WEC in all nonsingular regions.

Note also that, while naked singularities are a major feauture of our CTC-creating spacetimes, not all spacetimes containing naked singularities contain CTCs. For instance, there exist models for the collapse of dust-clouds which contain naked singularities, but no CTCs \cite{joshi}.
\newline

\noindent{\em{Outline.}}  In Section \ref{sec:wire}, we will focus on a particular version of the fast-light `wire' metric.  The version of the wire metric was chosen for simplicity of analysis:  Namely, it is easy to prove the arguments of Section \ref{sec:CTCs}.  It is axisymmetric, and meets the Null, Weak, and Strong Energy Conditions (SEC), but not the Dominant Energy Condition (DEC).  It is also infinitely long, therefore not asymptotically flat in one dimension.  However, metrics for finite-length wires and wires that meet the DEC are known to us, and are described in the Appendix.  All such wire metrics contain fast-light regions and meet the WEC.

 In Section \ref{sec:CTCs}, we demonstrate how a pair of fast-light wires can be used to create CTCs.  The arguments here are framed in terms of fast-light wires, but can be modified to apply to other spacetimes where `effective' superluminal travel is possible.
 
  In Section \ref{sec:outro}, we state our conclusions.
  
   In the Appendix, we briefly describe finite-length wires and wires that meet the DEC.  We also expand on our analysis of the WEC.

\end{section}  

\begin{section}{A Fast-Light Wire} \label{sec:wire}
\noindent{\em{The Metric.}}
We introduce an infinite, aximsymmetric `wire' metric in natural units, with

\begin{equation}
\mathcal{F} = \begin{cases}
1 + {\left(   \frac{1}{r} - \frac{1}{R}  \right)}^n &\text{if $r \leq R$}\\
1 &\text{else}
\end{cases}
\label{eq:F}
\end{equation}
where $R$ is an arbitrary positive constant, and $ n \geq 2 $.  The line element is
\begin{equation}
ds^2 = -\mathcal{F}\,dt^2 + \frac{1}{\mathcal{F}}\,dr^2 + dz^2 + r^2\,d\phi^2
  \label{eq:metric}
\end{equation}
This metric represents a positive mass-energy distribution such as that shown in Figure \ref{fig:SWsegment}.  It contains a central line singularity at $r=0$.  Because the factor of $\nicefrac{1}{\mathcal{F}}$ next to $dr^2$ never becomes infinite, this metric has no horizon.  This metric has a Lorentzian signature at all points, and is asymptotically flat.  It is a vacuum metric when $r > R$.  The boundary at $r=R$ is $C^{n-1}$ smooth:  That is, up to the $(n-1)^{th}$ derivative, both cases of $\mathcal{F}$ have zero-valued deriatives at $r=R$.  This is because all terms of the $(n-1)^{th}$ derivative of $\mathcal{F}$ will contain a factor of $\left(\frac{1}{r} - \frac{1}{R}\right)$. The requirement that $ n \geq 2 $ is the requirement that the patch be continuous to at least the first derivative.  Greater smoothness can be obtained with greater values of $n$. If $n \geq 3$, there is neither mass nor pressure at the boundary $r=R$.

\begin{figure}[ht]
  \centering
  \includegraphics[scale=0.25]{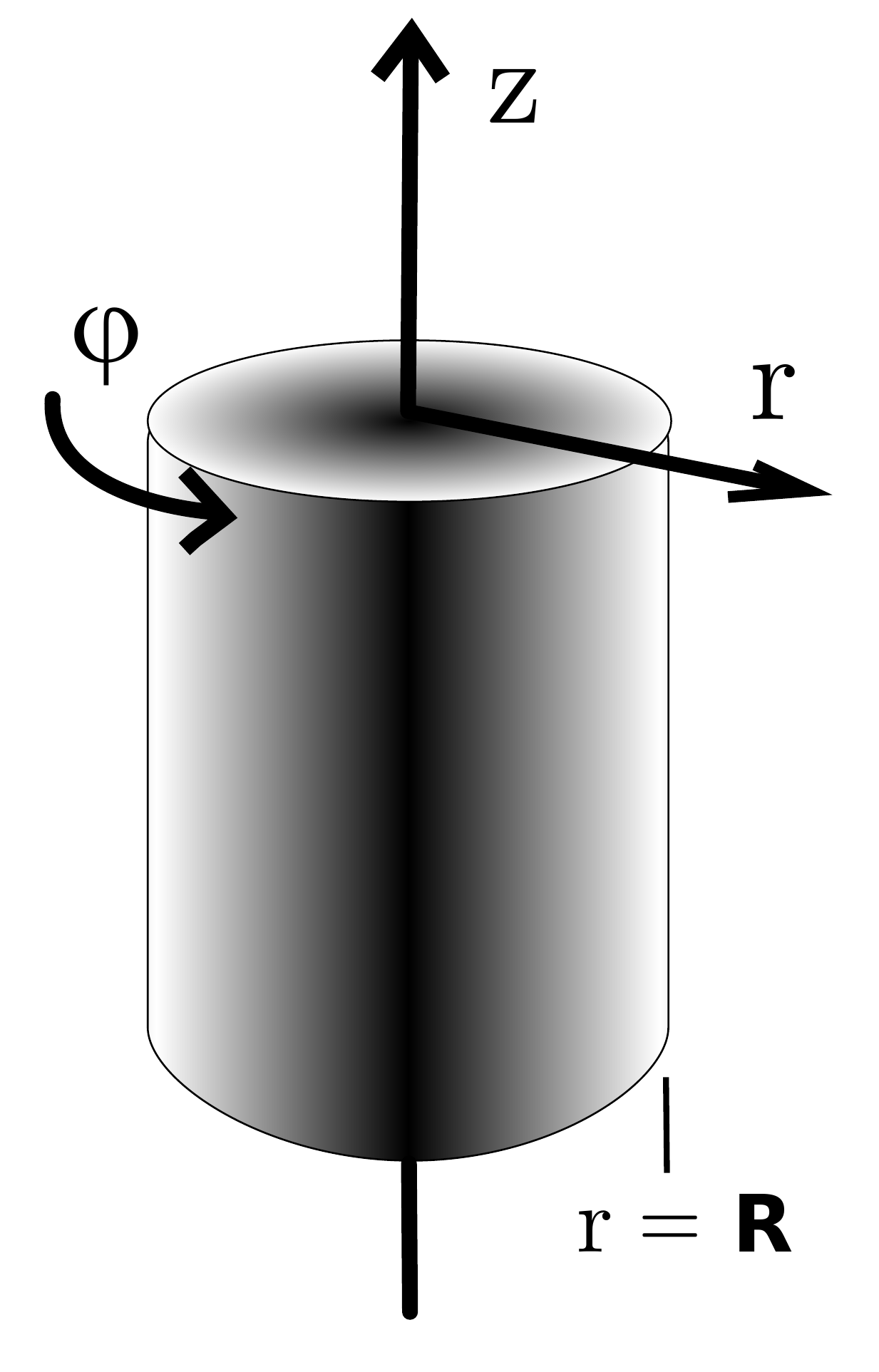}
  \caption{A segment of a fast-light wire, with the  dense central region colored dark. The wire has zero density at the boundary $r = R$, and it is surrounded by a flat vacuum spacetime. The wire density increases monotonically as $r\rightarrow0$.  The rate of increase is determined by $n$.}
  \label{fig:SWsegment}
\end{figure}

Note that we could have omitted the $-\frac{1}{R}$ term in (\ref{eq:F}), and simply had $\mathcal{F} = 1 + \frac{1}{r^n}$ for all $r$.  If we had done this, we would still obtain a wire metric with many of the same properties as we will describe below.  By including the $-\frac{1}{R}$ term, we ensure that our wire is of finite radial extent:  That is, beyond $r = R$, the wire exerts no gravitational influence on the surrounding space.  This greatly simplifies the analysis of CTCs in Section \ref{sec:CTCs}. 

To verify that the mass density described by the metric (\ref{eq:metric}) is indeed positive for any timelike observer, we must:
\begin{enumerate}
	 \item Compute the stress-energy tensor $T_{\mu\nu}$.  We use a zero cosmological constant.
	 \item  Find the transformation which orthonormalizes the metric $[\mathbf{g}]$, i.e. find the matrix $\Lambda$ such that $\Lambda^{T}[\mathbf{g}]\Lambda = [\mathbf{\eta}]$, where $[\mathbf{\eta}]$ is the Minkowski metric.  In index notation, we would write this as $\left(\Lambda^{T}\right)_{\rho}^{\mu} \mathbf{g}_{\mu\nu} \Lambda^{\nu}_{\sigma} = \mathbf{\eta}_{\rho \sigma}$.
	 \item  Perform this same transformation on the stress-energy tesnsor, i.e.:  $\left(\Lambda^{T}\right)_{\rho}^{\mu} T_{\mu\nu} \Lambda^{\nu}_{\sigma} = T_{ \hat{\rho}\hat{\sigma} }$, where our answer $T_{ \hat{\rho}\hat{\sigma} }$ is the orthonormalized stress-energy tensor. In other words, $T_{ \hat{\rho}\hat{\sigma} }$ is the stress-energy tensor as viewed by some local observer, to whom the space appears to be locally flat and the coordinate speed of light in natural units is always 1.  Orthonormalization simplifies analysis of the energy conditions in the space.
	 \item Test the energy conditions on this $T_{ \hat{\rho}\hat{\sigma} }$, to see if the metric (\ref{eq:metric}) exhibits energetically unrealistic behavior.  Energetically realistic spacetimes should, at minimum, satisfy the the Averaged Null Energy Condition (ANEC); the Weak Energy Condition (WEC) is somewhat more stringent.  In this paper we are primarily concerned with whether our metrics satisfy WEC at every nonsingular point in the spacetime.  See \cite{curiel} and the Appendix for details of how to determine if a metric meets the WEC.
\end{enumerate}

From ($\ref{eq:metric}$), we find that the unorthonormalized stress-energy tensor $T_{\mu\nu}$ in the range $0 < r \leq R$ is:

\begin{multline}
	T_{\mu\nu} = \\
	\mathcal{K} \times
	\left(\begin{array}{cccc} \mathcal{F}\left(R-r\right) & 0 & 0 & 0\\ 0 & \frac{-\left(R-r\right)}{\mathcal{F}} & 0 & 0\\ 0 & 0 & R\,\left(n-1\right) & 0\\ 0 & 0 & 0 & r^2\left(R-2\,r+R\,n\right) \end{array}\right)
\end{multline}
where
\begin{equation}
\mathcal{K} = \frac{R\,n}{16\pi}\frac{{\left(R-r\right)}^{n-2}}{{r^2\,\left(R\,r\right)}^n}
\end{equation}
The transformation matrix $\Lambda$ is given by:
\begin{equation}
	\Lambda = \left(\begin{array}{cccc} \frac{1}{\sqrt{\mathcal{F}}} & 0 & 0 & 0\\ 0 & \sqrt{\mathcal{F}} & 0 & 0\\ 0 & 0 & 1 & 0\\ 0 & 0 & 0 & \frac{1}{r} \end{array}\right)
\end{equation}
We use $\Lambda$ to obtain the orthonormalized stress-energy tensor:
\begin{multline}
	T_{ \hat{\rho}\hat{\sigma} } = \\
	\mathcal{K}	\times
	\left(\begin{array}{cccc} \left(R-r\right) & 0 & 0 & 0\\ 0 & -\left(R-r\right) & 0 & 0\\ 0 & 0 & R\left(n-1\right) & 0\\ 0 & 0 & 0 & \left(R-2\,r+R\,n\right) \end{array}\right)
	\label{eq:T}
\end{multline}

Since $0 < r \leq R$ and $n \geq 2$, all the terms in $T_{ \hat{\rho}\hat{\sigma} }$ are positive or zero, except $T_{ \hat{r}\hat{r} }$, the radial pressure.  $T_{ \hat{r}\hat{r} }$ is simply the negative of $T_{ \hat{t}\hat{t} }$, the mass-energy density of the wire. For a diagonal stress-energy tensor, the WEC requires that
\begin{equation}
	\begin{array}{r c l r}
		T_{ \hat{t}\hat{t} } &\geq& 0 &  
		\\
		T_{ \hat{t}\hat{t} } + T_{ \hat{i}\hat{i} } &\geq& 0 $\,\,\,\,for all spatial directions $\hat{i}
	\end{array}
	\label{eq:WECsimple}
\end{equation}

We have shown that the wire metric satisfies both of the conditions in ($\ref{eq:WECsimple}$) for $0 < r \leq R$.  When $R < r$, (\ref{eq:metric}) is simply vacuum, so it also satisfies these conditions.  Therefore, this wire satisfies the Weak Energy Condition everywhere.\footnote{$T_{ \hat{\rho}\hat{\sigma} }$ is technically undefined at the $r=0$ singularity itself, but meets the WEC for all arbitrarily small $r$.} For details on this form of the WEC and its use with an orthonormalized stress-energy tensor, see the Appendix.

Note that this wire also satisfies the Strong Energy Condition (SEC), which additionally requires that the trace of $ T_{ \hat{\mu}\hat{\nu} } $ is nonnegative \cite{curiel}.  However, this wire fails the Dominant Energy Condition (DEC), because the angular pressure $ T_{ \hat{\phi}\hat{\phi} } $ is greater in magnitude than the mass-energy density $T_{ \hat{t}\hat{t} }$. For a version of the wire that satisfies the DEC, see the Appendix.  (We do not consider the DEC-satisfying wire here, since it makes certain proofs significantly more complicated.  It may be of future interest to prove whether or not DEC-satisfying wires can support the CTCs which will be described in Section \ref{sec:CTCs}.)

The large negative radial pressure of (\ref{eq:WECsimple}) is notable, and may have some as-yet unclear role in CTC-creation.  However, this pressure is not negative enough to violate the WEC.  Therefore, the WEC is not an adequate way to rule out certain objectionable features of this spacetime, which are described in the next section. 

\hfill\newline

\noindent{\em{Features of the Wire Spacetime.}}
The wire metric is a `fast-light' metric in both the $r$ and $z$ directions.  In the case of the $z$ direction, the coordinate speed of light $c_z$ is greater than 1 because the metric's $\mathbf{g}_{tt}$ component has a magnitude greater than 1.  Note that this is opposite from the usual effect of a massive body on the spacetime around it.  For instance, the metric of a standard star can be considered a `slow-light' metric according to our definitions.

The wire metric bears some qualitative similarity to the `negative-mass Schwarzschild' metric, $ds^2 = -(1+\frac{2\abs{M}}{r})dt^2 + (1+\frac{2\abs{M}}{r})^{-1}dr^2 + r^2d\Omega^2$.  The `negative-mass Schwarzschild' is a fast-light metric, since $1+\frac{2\abs{M}}{r} > 1$.  Despite its name, the negative-mass Schwarzschild is a vacuum metric; i.e. it has an all-zero stress-energy tensor.  Its negative mass is implied to exist in the central singularity, while its explicit mass-energy density is zero.  By contrast, our wire metric has an explicitly positive mass-energy density for all timelike observers. 

During our investigation of the wire metric, we attempted to `patch over' the singularity at $r = 0$ using both analytic and numerical methods.  In other words, we tried to change the form of the metric as $r \rightarrow 0$ while keeping the metric the same at higher $r$.  We were unable to find any `patch' for the wire that could both remove the central singularity and preserve WEC at every point in the patch.  However, we could not prove generally that such a patch does or does not exist. Olum's Superluminal Condition \cite{olum} seems to imply that such a patch does not exist.

In the case of the wire, having $|\mathbf{g}_{tt}| > 1$ when $r \leq R$ makes the wire's gravity repulsive in that region.  This can be seen by calculating the geodesics of the wire metric, which we do below.

We start by taking the scalar product of an unknown timelike geodesic for a massive particle $(\dot{t},\dot{r}, \dot{z},\dot{\phi})^T$, where the dotted coordinates refer to derivatives with respect to an affine parameter.  Then, using the metric given by (\ref{eq:F}) and (\ref{eq:metric}), we have:
\begin{equation} \label{eq:rawmassive}
\left(\mathbf{g}_{\mu\nu}u^{\mu}u^{\nu}\right)_{massive}  = -1 = -\mathcal{F}\,\dot{t}^2 + \frac{1}{\mathcal{F}}\,\dot{r}^2 + \dot{z}^2 + r^2\dot{\phi}^2
\end{equation}
for a massive particle, while for massless particles:
\begin{equation} \label{eq:rawmassless}
\left(\mathbf{g}_{\mu\nu}u^{\mu}u^{\nu}\right)_{null} = 0 = -\mathcal{F}\,\dot{t}^2 + \frac{1}{\mathcal{F}}\,\dot{r}^2 + \dot{z}^2 + r^2\dot{\phi}^2
\end{equation} 
We will use the Euler-Lagrange equation $\frac{\partial \mathcal{L}}{\partial x^{\nu}} = \frac{d}{d\tau}\left( \frac{\partial \mathcal{L}}{\partial \dot{x}^{\nu}}\right)$ in combination with (\ref{eq:rawmassive}) and (\ref{eq:rawmassless}) to obtain the geodesics. Let
\begin{equation} \label{eq:Lagrangian}
\mathcal{L} = \frac{1}{2}\left( -\mathcal{F}\,\dot{t}^2 + \frac{1}{\mathcal{F}}\,\dot{r}^2 + \dot{z}^2 + r^2\dot{\phi}^2\right)
\end{equation}

Using standard simplification methods, we find:
\begin{equation} \label{eq:eos1}
	\begin{array}{ccc} 
	\dot{t} = \frac{E}{\mathcal{F}},  &  \dot{z} = \mathcal{Z},  &  \dot{\phi} = \frac{J}{r^2} 
	\end{array}
\end{equation}
where $E$ is the test particle's conserved initial energy density, $Z$ is the test particle's conserved momentum density in the $z$ direction, and $J$ is a conserved angular momentum density.  (That is, $Z$ and $J$ are quantities that the test particle brings with it when it approaches the wire from $r = \infty$.  They are free inputs to the equation for $\ddot{r}$.)  For the radial geodesics at $r \leq R$, we solve the Euler-Lagrange equation and obtain:
\begin{equation} \label{eq:r2dot_generic}
\ddot{r} =  \frac{n}{2{r^2}}\left(\frac{1}{r} - \frac{1}{R}\right)^{n-1}\times\frac{\left(E^2 - \dot{r}^2\right)}{\mathcal{F}} \,\, + \,\, \mathcal{F}\frac{J^2}{r^3}
\end{equation}
$\ddot{r}$ refers to the geodesic 4-acceleration of the test particle in the $r$ direction.  We set (\ref{eq:r2dot_generic}) equal to \ref{eq:rawmassive}) and (\ref{eq:rawmassless}) 
, and arrive at:

\begin{equation}
\label{eq:r2dot_both}
\begin{array}{l}
\ddot{r}_{massive} = \frac{n}{2{r^2}}\left(\frac{1}{r} - \frac{1}{R}\right)^{n-1} \left(1 + \mathcal{Z}^2 + \frac{J^2}{r^2}\right) \,\, + \,\,\mathcal{F}\frac{J^2}{r^3}
\\
\ddot{r}_{massless} = \frac{n}{2{r^2}}\left(\frac{1}{r} - \frac{1}{R}\right)^{n-1} \left(0 + \mathcal{Z}^2 + \frac{J^2}{r^2}\right) \,\, + \,\, \mathcal{F}\frac{J^2}{r^3}
\end{array}
\end{equation}
\par

It is easy to see by inspection that $\ddot{r}_{massive}$ must always be a positive number.  This means that a massive particle near the wire is spontaneously accelerated to higher $r$, according to the particle's proper time.  Therefore, the wire's gravity is repulsive to massive particles.

For massless particles, $\ddot{r}$ is also positive if $\mathcal{Z}$ or $J$ is nonzero.  This means that photons are deflected outward from the wire if they have any initial momentum in the $z$ or $\phi$ direction.  However, a photon aimed straight at the wire ($\mathcal{Z} = J = 0$) will not be repelled.

Although this repulsive gravity is unusual, it is not technically forbidden. A greater constraint on the wire's realism comes if it can be proven that the wire's central singularity cannot be `patched over' in a satisfactory way.  This constraint is only partially due to the fact that the wire is a naked singularity:  As currently formulated, the total mass-per-unit-length of this unpatched wire is actually infinite, which makes it unrealistic even if Cosmic Censorship is incorrect.  The infinite mass can be seen by integrating the mass energy density $T_{\hat{t}\hat{t}}$ between $R$ and some arbitrary lower limit $r'$:

\begin{equation}
\frac{mass}{length} =\frac{1}{8\pi} \int_{r'}^{R}\frac{Rn}{r^2}\frac{(R-r)^{n + 1}}{(Rr)^n} \sqrt{-\mathbf{g}}\,\,2\pi dr 
\end{equation}

where $\mathbf{g}$ refers to the metric determinant, and $\sqrt{-\mathbf{g}}$ is the standard volume element for metric integrations.  (For this wire metric, $ \sqrt{-\mathbf{g}} = r$.) This leads to the conclusion that

\begin{equation}
\lim_{r' \to 0} \,\left(\frac{mass}{length}\right) \,\,=\,\,\infty
\end{equation}

If the wire's center cannot be `patched over' to avoid infinite positive mass, this infinity would give us an additional reason that the wire is not a physically reasonable spacetime.  Pedagogically, it is an example of how spacetimes with exotic features may fail realism tests other than the energy conditions.

We detail the construction of the time machine in the next section.

\end{section}  

\begin{section}{Closed Timelike Curves (CTCs)}  \label{sec:CTCs}
In this section we demonstrate how a pair of parallel fast-light wires form CTCs when one wire is boosted in the z-direction.  CTCs are also known as `time machines' because they allow an observer moving along them to travel into his own past.
\par
The construction of the time machine is shown in Figure \ref{fig:timeMachine}. Two fast-light wires are set parallel to each other a trivial distance $d$ apart in the $x$ direction.  Here, `trivial' means that $d << L$, but we should have $d > 2R$, so that the wires do not interact gravitationally, and the space between them is perfectly flat.  This lack of gravitational interaction between the wires greatly simplifies CTC analysis.\par
\begin{figure}[ht]
  \centering
  \includegraphics[scale=0.35]{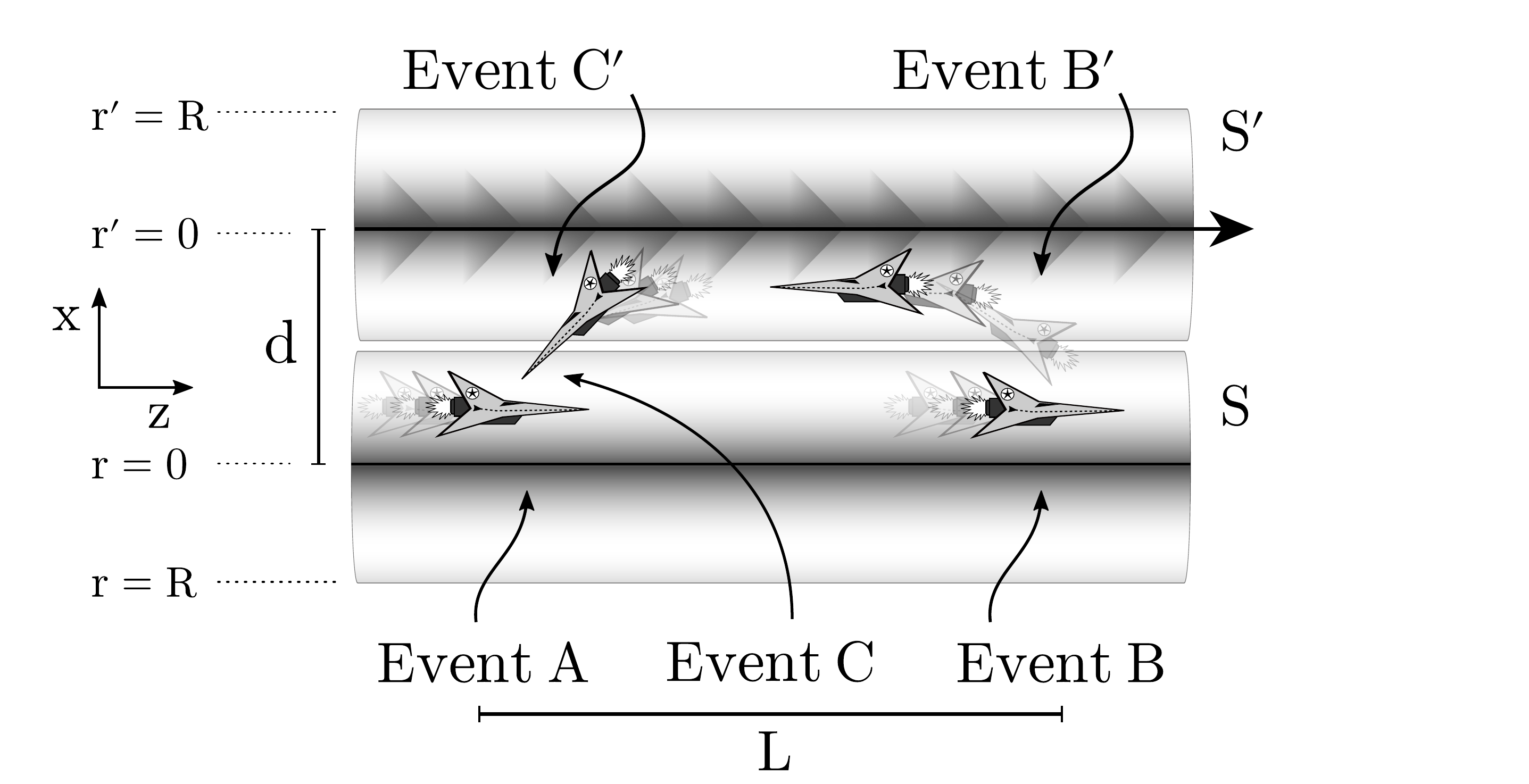}
  \caption{Two fast-light wires can be used to create a CTC.  Here, the top wire is boosted relative to the bottom wire, and a relativistic rocket follows a looped path between them.  If the rocket and the top wire are sufficiently fast, Event C occurs before Event A in frame S.  An animated version of this figure is available on YouTube: \url{https://youtu.be/ub6PGaygVwA}.}
  \label{fig:timeMachine}
\end{figure}
  We observe the wires from frame $S$, where the bottom wire is at rest.  Our coordinates represent the space as seen by observers (us) who are very far from the wires.  We will use natural units, where the standard speed of light $c$ in flat space is defined by $c = 1$.  The top wire is boosted by an amount $\beta < 1$ in the $z$ direction.  We say that the top wire is at rest in frame $S'$.  A rocket capable of relativistic velocities travels along the bottom wire from Event A to Event B (Figure \ref{fig:timeMachine}). The rocket will then turn around and travel back along the top wire to Event C, which has the same spatial location as Event A.  We will show that it is possible for Event C to occur before Event A in frame $S$.
\par
An animated version of Figure \ref{fig:timeMachine}, shown entirely from the point of view of an observer in the $S$ frame, is available on YouTube: \url{https://youtu.be/ub6PGaygVwA}.
\par
The time travel procedure is as follows:\newline

\noindent{\em{Step I.}}  The rocket starts at Event A.  In the $S$ frame, Event A has coordinates $(t, z) = (0,0)$.  For simplicity, we assume that the rocket is capable of travelling at very nearly the speed of light in its local space.  However, the rocket is not in flat space like us.  Because the rocket is in a fast-light region close to the wire, the coordinate speed of light at the rocket's location is greater than 1 (see (\ref{eq:c})).  Simply put, the rocket appears to be going faster than light.  Let's say that the coordinate speed of light at the rocket's location is $k > 1$, where the exact value of $k$ depends on how close the rocket is to the wire.\newline

\noindent{\em{Step II.}}
We can use this information to find the coordinates of Event B in $S$.  From Figure \ref{fig:timeMachine}, we can see that Event B is a distance of $L$ from Event A, so $\Delta z_{AB} = L$.  If the rocket can travel arbitrarily close to the coordinate speed of light, then it takes a time $\Delta t_{AB} = \frac{\Delta z_{AB}}{k} = \frac{L}{k}$ to get to Event B from Event A.  So Event B has coordinates $(t,z) = (\frac{L}{k}, L)$ in frame $S$.\newline

\noindent{\em{Step III.}} We'd like to find the coordinates of Event B in the $S'$ frame.  The $S'$ frame is simply the frame of a distant observer keeping pace with the top wire, so we do not need to consider the modifications to special relativity that would be necessary for an observer who was in the curved spacetime close to either wire.  That means a basic Lorenz boost will work:
\begin{equation} \label{eq:boostAB}
	\begin{split}
		\Delta t'_{AB} = \gamma(\Delta t_{AB} - \beta\Delta z_{AB}) = \gamma L\left(\nicefrac{1}{k} - \beta\right) \\
		\Delta z'_{AB} = \gamma(\Delta z_{AB} - \beta\Delta t_{AB}) = \gamma L\left(1 - \nicefrac{\beta}{k}\right)
	\end{split}
\end{equation}
Note that if $\beta > \nicefrac{1}{k}$ then Event B occurs earlier than Event A in frame $S'$, because the $\Delta t'_{AB}$ is then negative.\newline

\noindent{\em{Step IV.}}  From Event B, the rocket will cross to the top wire.  It arrives at Event B'. The distance of space the rocket must cross is $\Delta x$. $\Delta x$ should be very small relative to the length $L$, and somewhat smaller than the distance $d$ between the wires: That is, $\Delta x < d << L$.

   This requirement is important, because as long as the crossing time $\Delta t_{BB'}$ is small relative to $\Delta t_{AB}$, we can neglect it.  (Note that both these time intervals are as measured by the coordinate time in frame $S$.)  We saw in Step II that $\Delta t_{AB}$ is directly proportional to $L$.  The crossing time, on the other hand, should depend on $\Delta x$, and not dependent on $L$ at all.  So we ought to be able to ensure $\Delta t_{BB'} << \Delta t_{AB}$ by choosing a large enough $L$.  The only way this choice fails is if the crossing time in $S$ somehow becomes infinite.  This would only happen if the coordinate speed of light in the radial direction ($c_r$) becomes 0 in the vicinity of one of the wires.  It can be verified that $c_r \neq 0$.\footnote{ 
   For the wire which is at rest in the $S$ frame, we can use the wire metric (\ref{eq:metric}) and $c_r = \sqrt{|g_{tt}/g_{rr}|}$ (See Eq. \ref{eq:c}) to verify that $c_r \neq 0$ anywhere in the vicinity of the at-rest wire.  For the wire which is moving with velocity $\beta$ in the $z$-direction in the $S$ frame, we must first deboost this wire's metric into a coordinate system which is at rest in the $S$ frame.  We can do this using a tensor transformation.   We find that the $g_{tt}$ component of the deboosted metric is $g_{tt} = -\left(  1 + \gamma^2 \left(\frac{1}{r} - \frac{1}{R}\right)^n \right)$, $g_{rr} = \frac{1}{\mathcal{F}}$ (it is unaffected by a boost in $z$), while $g_{tr} = 0$.  Since $g_{tr} = 0$, we can use Eq. \ref{eq:c} to get $c_r$ here, too.  We find that $c_r$ cannot equal 0.}  Therefore, we can always choose an $L$ such that $\Delta t_{BB'} << \Delta t_{AB}$, and $t_{BB'}$ can be neglected.

There are several details to note here.  First, although the distance $d$ between the wires' centers should be negligibly small compared to $L$, we must have $d > 2R$, as stated earlier.  The gravitational effect of the wires ends a distance of $R$ away from them, so the $d > 2R$ requirement ensures the wires do not interact gravitationally.

Second, since the gravity of the wires is repulsive to massive objects, the rocket must exert more effort as it gets closer to the top wire.  We assume it is capable of doing this. 

Finally, it is important to note that we do not consider the proper time experienced by a passenger in the rocket.  The rocket accelerates intensely between $B$ and $B'$, and this would be relevant if we wanted to know how much the passenger ages relative to an observer in $S$.  However, we are ultimately concerned only with the passage of coordinate time in frame $S$:  Namely, is $\Delta t_{AC} < 0$, as measured in in $S$?  We do not need to consider the passenger in order to answer this question.  

At the end of Step IV, the rocket has reached the top wire at Event B', where B' $\approx$ B, by the logic above.\newline

\noindent{\em{Step V.}} The rocket travels back along the top wire from Event B' to Event C'.  This leg of the journey will have some $\Delta t'_{BC}$ and $\Delta z'_{BC}$ in the $S'$ frame.  $\Delta t'_{BC}$ may be deduced from the fact that the two wires are identical, and we can chose the rocket's return path to have the same $k$ as its outbound journey.  In that case, $\Delta t'_{BC} = -\frac{\Delta z'_{BC}}{k}$, where the negative sign reflects the fact that the return journey is in the $-z$ direction.  Note that since we are working in the $S'$ frame, where the top wire is at rest, we do not need to worry about special relativistic effects on $k$.

We can obtain an expression for $\Delta z'_{BC}$ by considering the fact that $\Delta z_{BC}$ must equal $-L$ in order for Event A and Event C to occur at the same spatial location. (By the same logic used Step IV, Event C $\approx$ Event C', where Event C occurs after the rocket has crossed back down to the bottom wire.)

  Using an inverse Lorenz boost and solving for $\Delta z'_{BC}$, we have:
\begin{equation} \label{eq:deboost_zBC}
	\begin{split}
		\Delta z_{BC} = -L = \gamma(\Delta z'_{BC} + \beta\Delta t'_{BC}) \\
		\Rightarrow \Delta z'_{BC}  = \frac{-L}{\gamma(1-\frac{\beta}{k})}
	\end{split}
\end{equation}

\noindent{\em{Step VI.}} We now have and expression for both $\Delta t'_{BC}$ and $\Delta z'_{BC}$ in the $S'$ frame.  Use them to get an expression for $\Delta t_{BC}$ in the $S$ frame, via another inverse Lorenz boost:
\begin{equation} \label{eq:deboost_tBC}
	\begin{split}
		\Delta t_{BC} = \gamma(\Delta t'_{BC} + \beta\Delta z'_{BC}) \\
		= \gamma\Delta z'_{BC} \left(-\frac{1}{k} + \beta\right) \\
		\Rightarrow \Delta t_{BC}  = \frac{L(\frac{1}{k} - \beta)}{(1-\frac{\beta}{k})}
	\end{split}
\end{equation}

\noindent{\em{Step VII.}}  Let's add $\Delta t_{AB}$ from Step II and $\Delta t_{BC}$ from Step VI together to get the total time difference between Events A and C in frame $S$:
\begin{equation} \label{eq:total_delta_t}
	\begin{split}
		\Delta t_{AC} = \Delta t_{AB} + \Delta t_{BC} 
		= \frac{L}{k - \beta}\left( 2 - k\beta - \frac{\beta}{k} \right)
	\end{split}
\end{equation}
It is quite possible for $\Delta t_{AC}$ to be a negative number.
\footnote{The $\Delta t_{AC}$ given in (\ref{eq:total_delta_t}) can be close to 0, in which case the $\Delta t_{BB'}$ which we neglected earlier may technically be relevant.  However, since the $\Delta t_{AC}$  of (\ref{eq:total_delta_t}) is proportional to $L$, and $\Delta t_{BB'}$ is not, we can generally ensure that $\Delta t_{BB'}$ is negligible by choosing sufficiently large $L$.}
  The factor of $\frac{L}{(k-\beta)}$ outside is always positive, so $\Delta t_{AC} < 0$ when
\begin{equation}\label{eq:ctc_condition}
\beta > \frac{2}{k+\frac{1}{k}} \,\Rightarrow\,\Delta t_{AC} < 0
\end{equation}
Equation (\ref{eq:ctc_condition}) is our CTC condition:  It is what makes Figure \ref{fig:timeMachine} a time machine. For instance, if $k = 3$ and $\beta > \nicefrac{3}{4
}$, then $\Delta t_{AC}$ is negative, and the rocket arrives on the left side of the figure before it ever set out.  For higher $k$ (which occur when the rocket is flying very close to the wires), the boosted wire does not need to be as fast in order to create the time travel effect; e.g. at $k = 10$ we only need $\beta > 0.198$.  In the limit $k\rightarrow\infty$, even nonrelativistic motions of the two wires create CTCs.\newline\par

It is of interest whether of not this spacetime is "totally vicious", as defined by Tipler \cite{tiplerVicious}.  That is, do CTCs pass through every point in the spacetime?  This is equivalent to asking whether or not a time traveler could access any event in the spacetime.  

In the case of infinite wires, the spacetime is indeed totally vicious: The time $\Delta t_{AC}$ given in Eq. (\ref{eq:total_delta_t}) is proportional to $L$, the length traveled along the wires, and $L$ may be infinite if the wires are infinite.  A time traveler could therefore travel arbitrarily far back in time. This would give the traveler access to events arbitrarily far from the wires:  For instance, to access an event that occured one year ago and a thousand light-years away from the wire, the traveler could simply use the wires to travel 1001 years into the past.  The total viciousness of the infinite-wire spacetime is a constraint on its realism.

A spacetime containing finite versions of the the wires is not totally vicious.  Two possible constructions of finite wires are given in the Appendix.  The argument for the existence of CTCs in such a spacetime is fundamentally similar to the arguments of Eqs. (\ref{eq:boostAB}) through (\ref{eq:ctc_condition}).  However, the evolution of the CTCs (if they exist) is significantly more complicated with finite wires:  This is because two finite wires moving relative to each other will only be in close proximity for a limited span of time.  Furthermore, the CTCs themselves would have limited scope: If we use Eq. (\ref{eq:total_delta_t}) as an approximation for the amount of time that can be "gained" by traveling along the finite wires, we see that it is proportional to $L$, which cannot be longer than the wires themselves.  This limitation holds even if our traveler travels arbitrarily close to the wires, where the coordinate speed of light $k$ becomes infinite:

\begin{equation}
\label{eq:limit_finite_wires}
\lim_{k\to\infty}\,\Delta t_{AC} = -L\beta
\end{equation} 

A time machine which uses finite wires can only travel a limited ways into the past; therefore a spacetime containing finite-wire CTCs is not totally vicious.  A full treatment of finite-wire CTCs and their implications is outside the scope of this work.\newline\par

As a final consideration, we might ask whether the CTC spacetime of Figure \ref{fig:timeMachine} really satisfies the WEC, since it contains two wires instead of one.  To answer this, consider that the WEC is a local condition:  Whether or not the WEC is met at a point depends only on the values of energy density and pressure at that point.  The wires in Figure \ref{fig:timeMachine} are of finite radial extent and are separated by flat Minkowski space (see Eq. \ref{eq:F} and \ref{eq:metric}): This means they do not interact gravitationally, and cannot affect each other's ability to meet the WEC. 

\end{section}  

\begin{section}{Conclusions}  \label{sec:outro}

We have demonstrated that the Weak Energy Condition (WEC) alone does not forbid the existence of `fast-light' regions in asymptotically flat spacetimes.  We have detailed the construction of one type of static `fast-light wire', and showed that two such wires can be used to create Closed Timelike Curves (CTCs).  We propose that the Weak Energy Condition is not an adequate standalone argument against superluminal travel.

\end{section}  

\begin{section}{Acknowledgements}
We would like to thank Amos Ori for helpful feedback on this paper. C.M. acknowledges support from the University of Massachusetts Dartmouth Graduate School. G.K. acknowledges research support from NSF Grants No. PHY-1414440 and No. PHY–1606333, and from the U.S. Air Force agreement No. 10-RI-CRADA-09. 
\end{section}

\begin{section}{Appendix}

We mentioned in the Outline that some of the wires' `realism' problems, except the singularity, can be addressed by modifying the metric given in (\ref{eq:metric}).  Here we briefly demonstrate.
\hfill\newline

\noindent{\em{A Wire that meets the Dominant Energy Condition.}} In the course of our investigation, we found versions of the fast-light wire which meet the dominant energy condition (DEC), in addition to the weak and strong energy conditions met by the basic wire.  We present one such wire here.  Let $\mathcal{H} = 1 + 2/r$, where $0 \leq r \leq \infty$ is the range of $r$.  Then our metric is
\begin{equation}
ds^2 = \mathcal{H}\,dt^2 + \frac{1}{\mathcal{H}}\,dr^2 + \frac{1}{\mathcal{H}}\,dz^2 + r^2\,d\phi^2
  \label{eq:DECmetric}
\end{equation}
By the same methods used in Section \ref{sec:wire}, we have:

\begin{equation}
	T_{ \hat{\mu}\hat{\nu} } = 
	\frac{1}{8 \pi r^3 (r+2)}
	\left(\begin{array}{cccc} 2\left(r + 1\right) & 0 & 0 & 0\\ 0 & -1 & 0 & 0\\ 0 & 0 & 0 & 0\\ 0 & 0 & 0 & 1 \end{array}\right)
	\label{eq:T_DEC}
\end{equation}
For a diagonal metric with a diagonal stress-energy tensor $T_{\hat{\mu} \hat{\nu}} \,=\, diag\left( \rho, P_1, P_2, P_3 \right)$, the DEC requires that $\rho > |P_i|$ for the mass-energy density $\rho$ and any pressure $P_i$. That condition is met here.  This wire also meets the Strong, Weak, and Null Energy Conditions at every point in the spacetime.

In terms of CTC creation, the primary problem with this wire is that it has infinite radial extent:  If a spacetime contains two of these DEC-preserving wires, they will gravitationally interact.  This complicates the CTC argument of Section \ref{sec:CTCs}, and possibly breaks it entirely, though there is as yet no proof whether or not a pair of DEC-preserving wires can support CTCs.  For this reason, we used the metric (\ref{eq:metric}) for most of this paper. It may be of future interest to prove whether or not preserving DEC destroys the CTCs. \newline

\noindent{\em{A Finite-Length Wire.}} It may also be of interest to know whether a fast-light wire can preserve WEC while being of finite length. There are two main ways to make a wire finite:  It must either be "capped" with two endpoints, or it must form a loop (Figure \ref{fig:patches}).  We briefly examine these possibilities.

\begin{figure}[ht]
  \centering
  \includegraphics[scale=0.2]{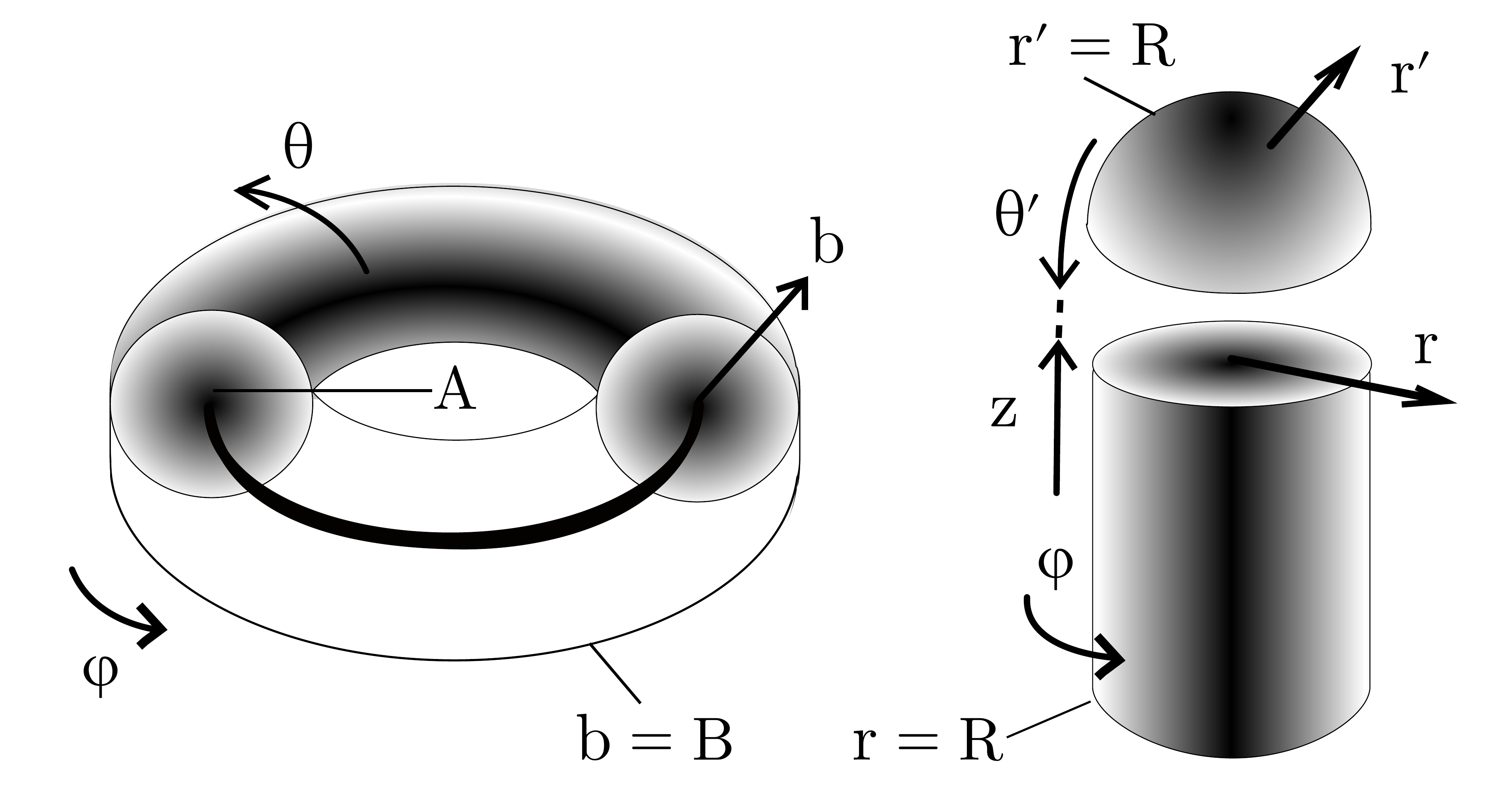}
  \caption{Two possible modifications to the wire metric to ensure finite length.  Left: Looping the wire.  This is done by mimicking the form of the wire metric (\ref{eq:metric}) on a toroidal coordinate system.  Right:  A segment of a wire with a hemispherical cap. The cap is made by mimicking the form of the wire metric on a spherical coordinate patch, and identifying $z$ with $\theta'$ at the boundary.}
  \label{fig:patches}
\end{figure}

\noindent ``Looping the wire'' (Figure \ref{fig:patches} left): This form of the wire is of particular interest, since ring singularities are thought to exist inside rotating black holes (such singularities, however, are not created the same way as a looped wire).  The wire can be looped by simply pasting the formulas of (\ref{eq:metric}) into torus-like coordinates.  Since most readers will be unfamiliar with the flatspace form of these coordinates, we present it here:

\begin{equation}
ds^2_{flat} = -dt^2 + db^2 + b^2d\theta^2 + (A - b\cos\theta)^2 d\phi^2	
\label{eq:flatspace}  
\end{equation}
where $A$ is the torus' fixed major radius, and $b$ is its minor radial coordinate, with $\theta$ and $\phi$ as shown in Figure \ref{fig:patches}.  Technically, a point in spacetime does not have a unique expression in this coordinate system.  Fortunately, that won't matter, because the non-vacuum portion of this metric is confined to the region $b < B < A$ (see Figure \ref{fig:patches} left).  Under these limitations, each non-vacuum spacetime point has a unique set of coordinates, and so our metric will be well-defined.  Let

\begin{equation}
\mathcal{Q} = 1 + {\left(\frac{1}{b} - \frac{1}{B}\right)}^n
\label{eq:wireTorusQ} 
\end{equation}
 where $n$ is an arbitrary number with $n \geq 2$, as before, and $B$ is the outer limit of the curved region of spacetime, so $0 \leq b \leq B$.  Then our looped-wire metric is given by
\begin{equation}
ds^2_{loop} = -\mathcal{Q} \,dt^2 + \frac{1}{\mathcal{Q}}\,db^2 + b^2d\theta^2 + (A - b\cos\theta)^2 d\phi^2	
\label{eq:wireTorus}  
\end{equation}
This metric is similar to (\ref{eq:metric}), with the $b$ coordinate taking the place of $r$. At $b = B$, this metric is patched into flat vacuum space in the same manner as the unlooped wire. The value of $n$ somewhat constrains the relation of $A$ and $B$, if we are interested in satisfying the WEC.  It is known that the orthonormalized stress-energy tensor for $n = 2$, $A = 2B$ has
\begin{equation}
\begin{array}{l}
   T_{\hat{t}\hat{t}} = \frac{1}{8\pi}\times \frac{\left(B-b\right)\,\left(2\,B^2-b\,\cos\theta\,\left(B+b\right)\right)}{B^2\,b^4\,\left(2\,B-b\,\cos\theta\right)}\\
   T_{\hat{b}\hat{b}} = -T_{\hat{t}\hat{t}} \\   
   T_{\hat{\theta}\hat{\theta}} = \frac{1}{8\pi}\times \frac{6\,B - 4\,b-b\,\cos\theta}{b^4\,\left(2\,B-b\,\cos\theta\right)}\\
   T_{\hat{\phi}\hat{\phi}} = \frac{1}{8\pi}\times\frac{1}{b^4}
\end{array}
\end{equation}
Since $b \leq B$ everywhere, and $-1 \leq \cos\theta \leq 1 $, all these values are positive or zero, except for the minor radial pressure $T_{\hat{b}\hat{b}}$, which has the same magnitude as the mass-energy density $T_{\hat{t}\hat{t}}$.  The sum of all these values is positive.  This `looped' version of the wire meets the NEC, WEC, and SEC while being of finite length.

One could imagine a chain of these loops which has the same net effect as a single straight wire.  In this way, it may be possible to use many loops to create the time machine described in Section \ref{sec:CTCs}. \newline

\noindent ``Capping the wire'' (Figure \ref{fig:patches} right): We make use of 2 hemispherical patches at arbitrary endpoints on the $z$-axis of the wire.  Supposing the caps are identical, we need only examine one of them.

To define this cap, we use spherical coordinates.  In the looped wire, we ported the formulas of (\ref{eq:metric}) into toroidal coordinates, and we do the same thing here to create the hemispherical cap.  The cap metric is then:
\begin{equation}
ds^2 = \mathcal{F}\,dt^2 + \frac{1}{\mathcal{F}}\,dr'^2 + r'^2\,d\theta'^2 + r'^2\,\sin^2\theta'\,d\phi'^2
  \label{eq:metricCap}
\end{equation}
where $\mathcal{F}$ is the same $\mathcal{F}$ given in (\ref{eq:F}), but with its cylindrical $r$ redefined as $r'$, a spherical coordinate.  This metric is valid in the range $0 \leq \theta' \leq \frac{\pi}{2}$.  At the surface $\theta' = \frac{\pi}{2}$ in the cap's coordinates, we patch the cap to the wire metric at some arbitrary surface of constant $z$ in the wire's coordinates.  The resulting stress-energy tensor for the cap has:
\begin{equation}
\begin{array}{l}
   T_{\hat{t}\hat{t}} = \frac{1}{8\pi}\times \frac{\left(r+R\,\left(n-1\right)\right)\,{\left(R-r\right)}^{n-1}}{R^n\,r^{n+2}} \\
   T_{\hat{r}\hat{r}} = -T_{\hat{t}\hat{t}} \\
   T_{\hat{\theta}\hat{\theta}} = \frac{1}{8\pi}\times \frac{n\,\left(n-1\right)\,{\left(R-r\right)}^{n-2}}{2R^{n-2}\,r^{n+2}}\\ 
  T_{\hat{\phi}\hat{\phi}} = T_{\hat{\theta}\hat{\theta}} \\     
\end{array}
\end{equation}
Primes such as $r'$ have been left off the coordinates here to avoid clutter.

Again, this meets the WEC.  However, these caps were not a main focus of research, and issues such as pressure discontinuities at the $\theta' = \frac{\pi}{2}$ patch boundary may become problematic.  (Note that the looped wire metric is much less likely to encounter these issues, because it is more continuous.)  We therefore present the caps only to demonstrate that `capped' solutions may exist.\newline

\noindent{\em{The Weak Energy Condition (WEC).}}
In its most general form, the WEC requires that:
\begin{equation}
	T_{ \mu\nu}u^{\mu}u^{\nu} \geq 0
	\label{eq:WEC}
\end{equation}
for every valid 4-velocity $u^{\mu}$ in the spacetime.  This condition is equivalent to saying that no timelike observer in the space can see negative mass.  We can simplify analysis if this requirement by dividing through by $(u^t)^2$ and rewriting the WEC as:
\begin{eqnarray}  
	T_{ tt } + 2T_{ ti }v^{i} + T_{ ij }v^{i}v^{j} & \geq & 0 
	\\
	\Rightarrow \,\rho + P_{ i }(v^{i})^2 & \geq & 0\, \text{for diagonal}\, T_{ \mu\nu }  
	\label{eq:WECo}
\end{eqnarray}

where the $v^{i}$ represent all possible timelike spatial 3-velocities.\par
In coordinates which are not orthonormal and a space which is not flat, it can be an enormous chore to test the WEC.  For one thing, each point in the space may have different constraints on what makes a vector `timelike'.   Orthonormalization removes that difficulty.  It describes the stress-energy tensor and the 3-velocities from the point of view of a local observer, to whom space is always locally flat, and the coordinate speed of light is the same constant value in every direction, i.e. $c = 1$. This means that, for all timelike vectors in orthonormal coordinates, $|v^{\hat{i}}| \leq 1$:  This makes it much easier to test the WEC. \par
Since the $v^{\hat{i}}$ in in orthonormalized version of ($\ref{eq:WECo}$) can take any value such that $|v^{\hat{i}}| \leq 1$, we should consider the $|v^{\hat{i}}| = 0$ case, corresponding to an observer at rest relative to the wire.  In this case, only the first term in ($\ref{eq:WECo}$) is nonzero, so we must have $T_{ \hat{t}\hat{t} } \geq 0$ for all valid $r$ in order to meet the WEC.  Looking at the form $T_{ \hat{t}\hat{t} }$ given in ($\ref{eq:T}$), we see that the wire metric meets this requirement: There is no way that $T_{ \hat{t}\hat{t} }$ can be negative for any $r$ in the range $0 \leq r \leq R$.  (Recall that at  $r > R$, we switch over to a vacuum metric, where the stress-energy tensor is zero and there is no possibility of violating WEC.)  \par

For all other possible observer velocities in this space, note that WEC cannot be violated if $T_{ \hat{t}\hat{t} } + T_{\hat{i}\hat{i}} \geq 0$ for every spatial direction $\hat{i}$.  This is because all velocities are quadratic sums of velocities in the 3 cardinal directions ($r, z, \phi$), and no velocity may have a magnitude greater than 1.  This is precisely the WEC requirement given in ($\ref{eq:WECsimple}$), which the wire metric fulfills.\par\par

\end{section}  

\end{document}